\documentclass[amsmath,amssymb,superscriptaddress,aps,a4paper,prl,reprint]{revtex4-1}

\usepackage{amssymb}
\usepackage{graphics} % Include figure files
\usepackage{lmodern}
\usepackage{textcomp}
\usepackage[T1]{fontenc} \usepackage{gensymb} %\micro, \ohm, \degree, \celsius
\usepackage[seperr, load={}]{siunitx}

\newcommand{\Article}{Letter}

\newcommand{\coloron}{(Color online)\ }
\newcommand{\figref}[1]{Fig.~\ref{#1}}
\newcommand{\dng}{\Delta n_g}

\begin{document}
\title{Single Cooper-pair pumping in the adiabatic limit and beyond}

\author{S. Gasparinetti}
	\email{simone.gasparinetti@aalto.fi}
	\affiliation{Low Temperature Laboratory, Aalto University, P.O. Box 15100, FI-00076 Aalto, Finland}
\author{P. Solinas}
	\affiliation{Low Temperature Laboratory, Aalto University, P.O. Box 15100, FI-00076 Aalto, Finland}
	\affiliation{Department of Applied Physics, Aalto University School of Science,
	P.O.Box 11100, FI-00076 Aalto, Finland}
\author{Y. Yoon}
	\affiliation{Low Temperature Laboratory, Aalto University, P.O. Box 15100, FI-00076 Aalto, Finland}
\author{J. P. Pekola}
	\affiliation{Low Temperature Laboratory, Aalto University, P.O. Box 15100, FI-00076 Aalto, Finland}

\begin{abstract}
We demonstrate controlled pumping of Cooper pairs down to the level of a single
pair per cycle, using an rf-driven Cooper-pair sluice. We also investigate the
breakdown of the adiabatic dynamics in two different ways. By transferring many
Cooper pairs at a time, we observe a crossover between pure Cooper-pair and
mixed Cooper-pair-quasiparticle transport. By tuning the Josephson coupling
that governs Cooper-pair tunneling, we characterize Landau-Zener transitions in
our device. Our data are quantitatively accounted for by a simple model including
decoherence effects.
\end{abstract}

\maketitle

Charge pumps \cite{Blumenthal2007,Giblin2010,Giazotto2011,Nevou2011} and
turnstiles \cite{Pekola2008a} have recently attracted considerable attention.
They could be used as building blocks for quantum computing devices
\cite{Feve2007}, or to create a quantized current source that would pair up with
the Josephson voltage and quantum Hall resistance to close the so-called quantum
metrology triangle \cite{Flowers2004}. Among different types of realizations, Cooper-pair pumps
\cite{Pekola1999,Niskanen2003,Leone2008a} stand out as macroscopically coherent
objects, with the phase of the superconducting order parameter in the leads playing a key role. In addition, the cyclic path
described in the space of parameters when pumping is equipped with a nontrivial
geometric structure, allowing for the observation of geometric-phase effects
\cite{Mottonen2006}.
In the adiabatic limit, a general relation was derived \cite{Aunola2003}
connecting the pumped charge to the geometric (Berry) phase accumulated by the system ground
state along a pumping cycle. This relation was experimentally demonstrated in
Ref.~\onlinecite{Mottonen2008}. Beyond the adiabatic limit, we have recently
proposed to employ a Cooper-pair pump as a Landau-Zener interferometer for geometric
phases \cite{Gasparinetti2011a}.

In this \Article, we demonstrate controlled pumping of a single Cooper pair,
using an rf-driven Cooper-pair sluice \cite{Niskanen2003,Niskanen2005}.
Accessing this regime opens new possibilites for Cooper-pair pumping, from
quantum metrology to the study of dissipation in driven quantum systems
\cite{Pekola2010}, also in connection with geometric phases.
We then investigate the breakdown of adiabatic pumping. In the sluice, this is
expected to take place via Landau-Zener transitions (LZTs) at level
anti-crossings.
We reach the nonadiabatic limit in two different ways. By pumping many Cooper pairs at a
time, we witness a crossover between pure Cooper-pair and mixed
Cooper-pair-quasiparticle dynamics, due to continuous generation of
nonequilibrium quasiparticles by the nonadiabatic drive.
By tuning the Josephson coupling that governs Cooper-pair tunnelling, we
characterize LZTs in our device. Our data are quantitatively accounted for by a
simple model comprising LZTs and realistic decoherence.

\begin{figure}
\center
\includegraphics{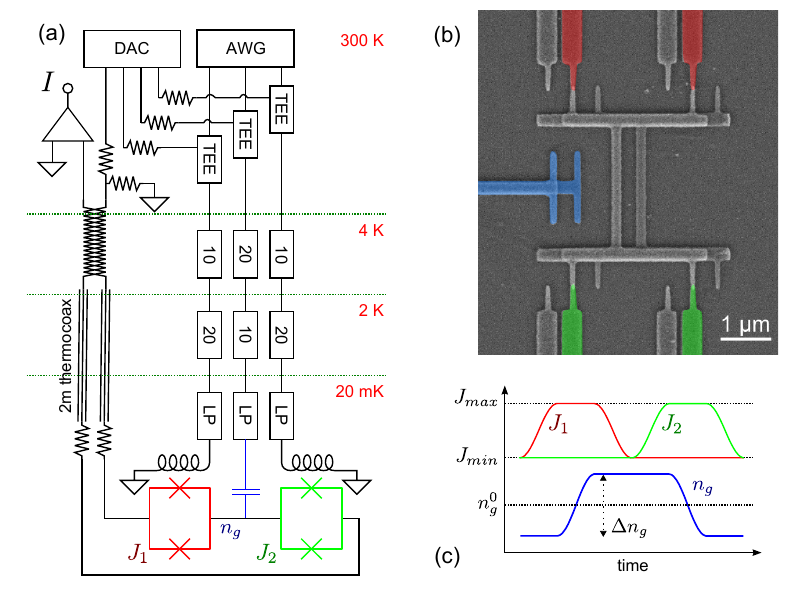}
\caption{\coloron The Cooper-pair sluice.
(a) Equivalent circuit of the sluice and scheme of the measurement set-up.
(b) False-color scanning-electron micrograph of a representative device.
(c) Time evolution of the control parameters for a typical pumping cycle.
% The gate is used as a piston to change the number of Cooper-pairs on the island,
% the SQUIDs as valves so as to impart direction to the flow of charge.
}
\label{fig:intro}
\end{figure}

\smallskip

The Cooper-pair sluice (\figref{fig:intro}) is a fully tunable Cooper-pair
transistor, consisting of a small superconducting island connected to leads by
two superconducting quantum interference devices (SQUIDs). The SQUIDs are
controlled independently by inductively coupled on-chip coils, so that they can
serve as Josephson junctions of tunable energy $J_1$, $J_2$. A gate electrode
capacitively coupled to the island induces a polarization charge $n_g=C_g
V_g/2e$ in units of Cooper pairs, where $V_g$ is the gate voltage and $C_g$ the
cross-capacitance between the gate and the island. The device is typically
operated in the charging regime, meaning that $4E_C \gg J_1,J_2$
($E_C=e^2/2C_\Sigma$ is the single-electron charging energy of the island,
$C_\Sigma$ being its total capacitance).
Pumping is realized by steering the three control parameters $J_1$, $J_2$ and
$n_g$ in a periodic fashion. The gate is used as a piston to change the number
of Cooper-pairs on the island, while the SQUIDs are operated as valves so as to
impart a direction to the flow of charge.
A typical pumping cycle is described in \figref{fig:intro}(c).

We fabricate the devices by standard electron-beam lithography, two-angle Al
evaporation, and liftoff. Small Josephson junctions (area $\approx$
70$\times$\SI{70}{nm}) are obtained by oxidization of the first Al layer in
controlled O$_2$ atmosphere.
A scanning-electron micrograph of a representative device is shown in
\figref{fig:intro}(b).
From the Coulomb-blockade conductance peak measured at \SI{2}{K} we obtain
$E_C=\SI{0.77}{K}$.
The normal-state resistance of the device at \SI{2}{K} is
$R_N=\SI{29}{k\Omega}$. Using the measured superconducting gap
at base temperature $\Delta=\SI{180}{\micro eV}$,
$R_N$ and the Ambegaokar-Baratoff formula, we estimate a maximum Josephson
energy $J^0_{max}=\SI{0.46}{K}$ per SQUID.
From switching statistics to the normal state in a current-biased configuration,
we estimate the ratio between the maximum and minimum Josephson couplings
obtained by varying the flux to be $J^0_{min}/J^0_{max} \leq 0.03$ for both
SQUIDs.

All measurements are performed in a dilution refrigerator down to 20 mK. The
set-up is schematically shown in \figref{fig:intro}(a). The SQUIDs and gate are
controlled by a combination of dc and rf signals, mixed together by bias tees.
The rf signals are generated by synchronized arbitrary waveform generators (AWG
in the Figure), guided to the sample by coaxial lines, attenuated and
thermalized at different temperature stages. Low-pass filters (LP) with
\SI{60}{dB} attenuation up to \SI{40}{GHz} are placed at the sample stage.
The dc wiring consists of \SI{160}{\Omega} surface-mount resistors,
\SI{2}{m}-long lossy coaxial lines, and \SI{1}{m}-long twisted pairs. At room
temperature, a voltage bias $V_b$ produced by a floating digital-to-analog
converter (DAC) is applied through a divider, and current is read out by a
transconductance amplifier with sensitivity \SI{e-10}{A/V}.
The sample is protected by two nested rf-tight shields in order to prevent
microwave irradiation from higher-temperature stages.

\begin{figure}
\center
\includegraphics{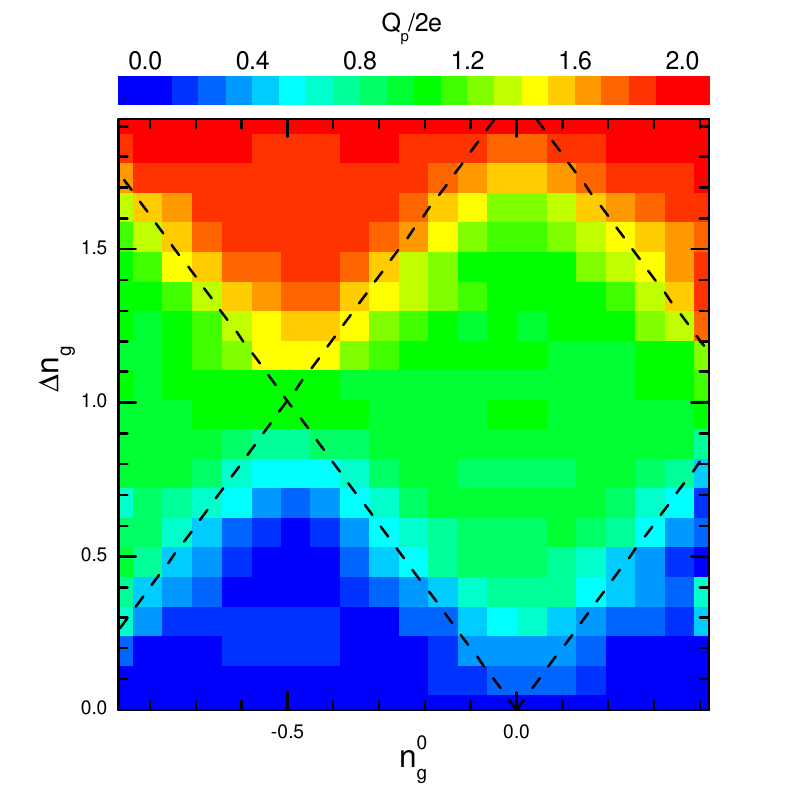}
\caption{\coloron
%(a) Representative radiofrequency pulses applied to the control parameters in order to achieve pumping.
Single Cooper-pair pumping. Pumped charge $Q_p$ versus peak-to-peak amplitude
$\dng$ and offset $n_g^0$ of the gate drive.
Dashed lines enclose the diamond-shaped regions where $Q_p/2e$ is expected to
be constant and quantized (assuming ideal operation).}
\label{fig:pumping}
\end{figure}

We measure the sluice in the supercurrent branch, close to $V_b=0$.
We apply control pulses at frequency $f$ ($f=\SI{80}{MHz}$ unless otherwise
stated). $V_b$ is nulled by minimizing the current flowing without applying any
pulses, thus compensating for the voltage induced by the current amplifier.
Application of the flux pulses alone was found not to shift the
zero-bias point, nor induce any additional current. This rules out the presence of rectification
effects, which could have been introduced, e.~g., by capacitive couplings
between the on-chip coils and the leads.
To further reduce the influence of voltage fluctuations, the pumped current
$I_p$ is detected as the difference between the currents measured while pumping
in opposite directions.

\smallskip

Evidence of single Cooper-pair pumping is presented in \figref{fig:pumping},
where the pumped charge per cycle $Q_p=I_p/f$ is plotted versus the offset
$n_g^0$ and the peak-to-peak amplitude $\dng$ of the gate drive. Since the
pumping cycle of \figref{fig:pumping} is fully adiabatic, we expect $Q_p$ to be
quantized in units of Cooper pairs, the first correction being of the order of
$J_{min}/J_{max}$ \cite{Mottonen2006}. From energy-diagram considerations, it is
easy to show that the regions of constant $Q_p$ are diamond-shaped in the
offset-amplitude plane. The regions are delimited by the family of curves
$\dng=2|n_g^0+m|$, where $m$ can be any integer.

A remarkable feature of
\figref{fig:pumping} is that $Q_p$ is $2e$-periodic in the gate charge (that is,
the size of the diamonds is 1 unit along the $n_g^0$ and 2 units along the
$\dng$ axis).
Previous measurements with the Cooper-pair sluice
\cite{Niskanen2005,Vartiainen2007} reported $1e$-periodic plateaus in the pumped
charge plotted against the amplitude of the gate drive (no dependence on the
gate offset was reported).
The authors ascribed the observed periodicity to quasiparticle poisoning.
Quasiparticle poisoning \cite{Joyez1994} has been intensively studied in systems
closely related to the sluice, the single Cooper-pair transistor
\cite{Aumentado2004,Naaman2006,Ferguson2006} and the single Cooper-pair box
\cite{Shaw2008,Persson2010}. All these devices feature a superconducting island
in the Coulomb-blockade regime. For a given position of the gate, there are two
metastable states for the island (``odd state'' and ``even state''), differing
by the presence of one quasiparticle. Nonequilibrium quasiparticles generated in
the leads (at temperatures $T \ll \Delta/k_B$, the thermal population of
quasiparticle states is unimportant) drive transitions between the two states
(``parity fluctuations''), shifting the gate charge by exactly half a Cooper
pair.
In the sluice, this results in $1e$-periodic pumped charge, provided that the
time scale of parity fluctuations is intermediate between the pumping period and
the acquisition time \cite{Niskanen2005,Vartiainen2007}.
In our case, the clean $2e$ periodicity observed implies that the device is not
``poisoned'' by quasiparticles. We ascribe this improvement to two factors:
efficient microwave shielding and clean grounding of the probe leads.
The importance of microwave shielding has been emphasized in other recent
experimental works \cite{Barends2011,Saira2012}.

\begin{figure}
\center
\includegraphics{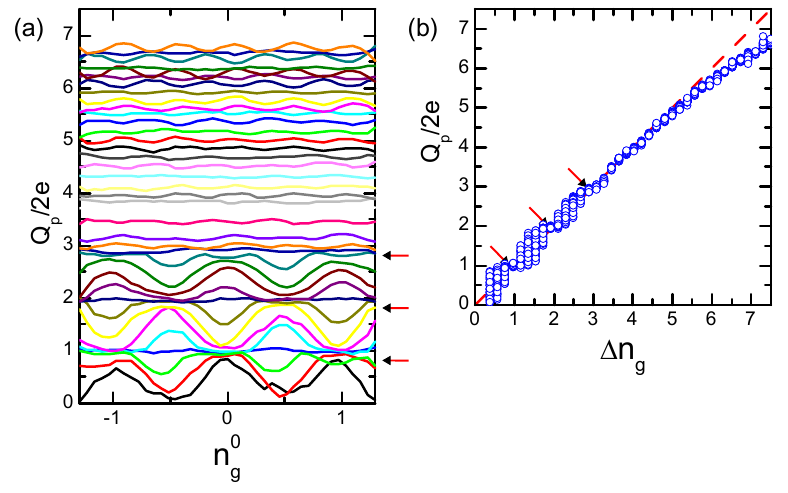}
\caption{\coloron Crossover between pure Cooper-pair and mixed
Cooper-pair-quasiparticle dynamics.
(a) Pumped charge $Q_p$ versus gate offset $n_g^0$
for increasing gate amplitudes $\dng$ (bottom to top).
(b) Circles: same data as in (a), projected on the $n_g^0$ axis and plotted
versus $\dng$. Dashed line: asymptotic adiabatic-limit expectation. In
both panels, the first three pumping plateaus are indicated by arrows.
}
\label{fig:quasip}
\end{figure}

While the data of \figref{fig:pumping} are clearly not affected by
nonequilibrium quasiparticles, the latter may come into play as a result of
nonadiabatic pumping. One way of making the pumping nonadiabatic is to increase
the amplitude of the gate modulation. This increases both the effective speed of
the drive (continuously), and the number of tunnelling events involved
(discretely).
We do it in \figref{fig:quasip}(a), where $Q_p$ is plotted versus $n_g^0$ with
$\dng$ taking a series of values in the range of $0.1$ and 7. The same data are
also plotted against $\dng$ in \figref{fig:quasip}(b). The data show a clear
crossover between pure Cooper-pair and mixed Cooper-pair-quasiparticle dynamics.
Up to about $\dng=3$, $Q_p$ is $2e$-periodic in $n_g^0$, as in
\figref{fig:pumping}. The pumping plateaus are also $2e$-periodic in $\dng$.
They show up as straight lines in \figref{fig:quasip}(a), and
nodes in \figref{fig:quasip}(b). The first three plateaus are indicated by
arrows.
The crossover takes place between about $\dng=3$ and $\dng=5$, where the pattern is blurred.
Finally, for $\dng \gtrsim 5$ a clear periodicity is restored, but the period
has doubled. These data show that quasiparticle poisoning, while initially
absent, can be induced by nonequilibrium quasiparticles generated by a
nonadiabatic drive.
The link between loss of adiabaticity and quasiparticle poisoning is
strengthened by the fact that the crossover
accompanied by a reduction in $Q_p$ with respect to the adiabatic-limit
expectation, as shown in \figref{fig:quasip}(b) for $\dng \gtrsim 5$. As
nonadiabatic transitions occur in the sluice as missed Cooper-pair tunnellings,
they leave a detectable trace in overall magnitude of the pumped charge.

We understand the generation of nonequilibrium quasiparticles as a multi-step
process: LZT to an excited state followed by relaxation via
Cooper-pair breaking and quasiparticle tunnelling.
When the drive is nonadiabatic, one or more Cooper-pairs may fail to tunnel as
dictated by the gate, leaving the island in an excited state.
Even if there are no quasiparticles in the leads near the junctions,
quasiparticle tunnelling from/into the island is still possible provided the
energy gain $\Delta E^\pm$ of the process \cite{note_qptunnel} exceeds the
energy cost $2\Delta$ required to break a Cooper pair. In our case, due to the moderate charging energy of our device
($E_C=0.33\Delta$) this is possible when the charge occupation of the
island differs from that of the ground state by at least two Cooper-pairs.
After the tunnelling of a quasiparticle, the island is left in a metastable
state with an odd quasiparticle number, which may then decay by tunnelling of a
second quasiparticle. Overall, this mechanism is similar to the well-known
Josephson-quasiparticle cycle \cite{Fulton1989}, with the nonadiabatic gate
drive playing the role of an effective voltage bias in dynamically creating
nonequilibrium.

\smallskip

\begin{figure}
\center
\includegraphics{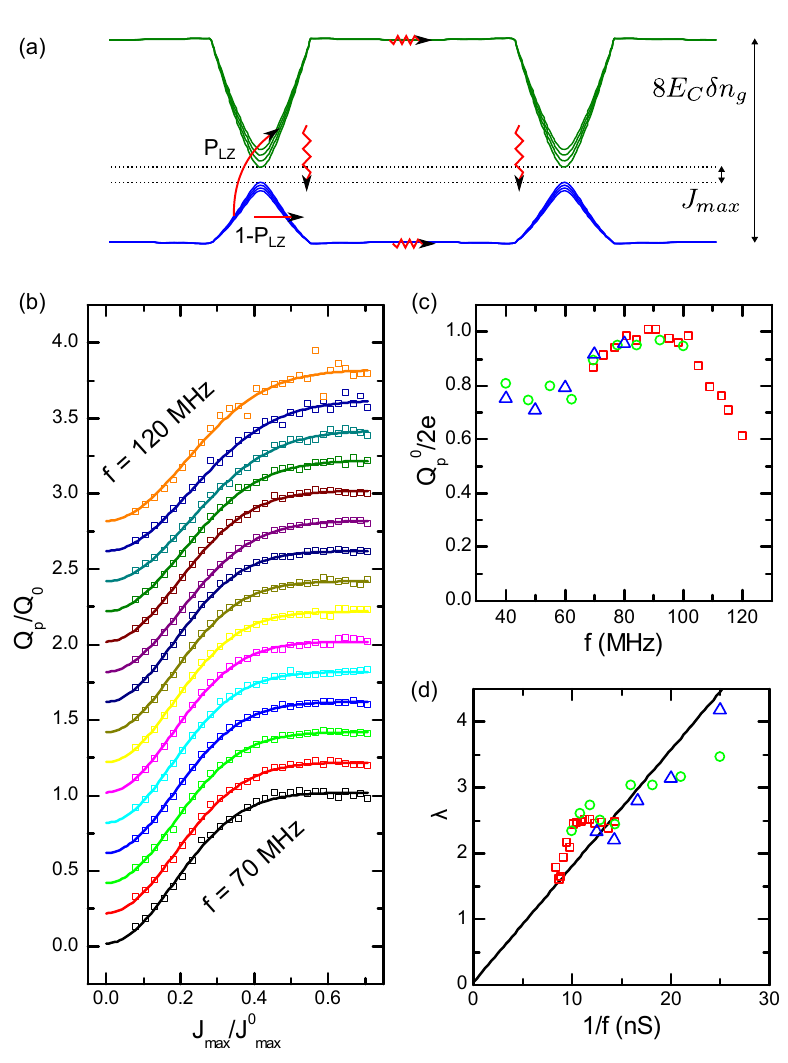}
\caption{\coloron
Nonadiabatic pumping and Landau-Zener transitions.
(a) Instantaneous energies of the ground and first
excited state versus time for a pumping period. In the model described in the
text, nonadiabatic transitions are localized at level crossings, and occur with
probability $P_{LZ}$. Decoherence induces complete
dephasing between subsequent transitions, and relaxation via inelastic
Cooper-pair tunnelling (both are indicated by wavy arrows).
(b) Normalized pumped charge $Q_p/Q_0$ versus Josephson coupling $J_{max}$, for
a set of frequencies in the range of 70 and 120 MHz.
The experimental traces (squares) are vertically offset by 0.2, and plotted
together with the best-fit of Eq.~\ref{eq:easyqp} to them (solid lines).
(c,d) Asymptotic pumped charge $Q_0$ versus $f$ (c) and $\lambda$ parameter
versus $1/f$ (d) for three different measurement sets (squares, circles, and
triangles). The solid line in (d) is a fit to the data of the expression
$\lambda=f_0/f$.}
\label{fig:nonadi}
\end{figure}

We now take a closer look at individual LZTs.
To do so, we choose to pump a single Cooper pair at a time. This reduces the
dynamics to that of a two-level system. Transitions between the adiabatic ground
($g$) and excited state ($e$) may occur due to Landau-Zener tunnelling at
avoided crossings [see \figref{fig:nonadi}(a)].
For a single crossing, the transition probability
$P_{LZ}=e^{-2\pi\delta}$ is governed by the adiabatic parameter
$\delta=J_{max}^2/\hbar v$, where $J_{max}$ is the Josephson coupling of the
active SQUID at the crossing and $v=E_C d n_g/dt$ the rate of change of the
energy difference between diabatic (charge) states. The full tunability of our
device gives us the possiblity to control the degree of adiabaticity in several
independent ways.
In \figref{fig:nonadi}(b), we plot $Q_p$ versus $J_{max}$, for the case
$n_g^0=0$ and $\dng=0.45$. The traces are taken at different pumping frequencies
in the range of 70 and \SI{120}{MHz}, normalized to the asymptotic pumped charge
$Q_0$, and vertically offset for clarity.
All traces approach a constant value for large values of $J_{max}$, and
monotonically decrease to $0$ for $J_{max}/J_{max}^0 \to 0$.

In order to understand the data quantitatively, we must consider the interplay
between LZTs and environment-induced decoherence. Decoherence effects in the
sluice have been studied theoretically with a master-equation approach
\cite{Pekola2010,Solinas2010,Russomanno2011}. Numerical calculations indicate
that even in the presence of modest decoherence, the driven Cooper-pair sluice
approaches a quasistationary state \cite{Russomanno2011} after a few tens
of cycles.
Since our acquisition time extends over about $10^6$ pumping cycles, we identify
$Q_p$ with the quasistationary charge pumped by the sluice in the
presence of its electromagnetic environment, averaged over a great number of
cycles.

Let us now discuss the role of decoherence in determining the quasistationary
state and hence $Q_p$.
In the present set-up, the sluice is directly connected to room-temperature
leads. Hence we expect a fast dephasing time, of the order of a few ns. This
implies that subsequent LZTs are totally uncorrelated, which rules out the
possibility of Landau-Zener-St\"uckelberg interference \cite{Shevchenko2010}.
The latter might instead play a role at higher pumping frequencies and/or in a
closed-loop geometry \cite{Gasparinetti2011a}.
Besides dephasing, we also have to consider relaxation, which can take place via
environment-assisted (inelastic) Cooper-pair tunnelling through the SQUIDs.
The energy scales and base temperature for our system are such that the
transition rates to the excited state are exponentially suppressed, so that the
environment is effectively at zero temperature. As a result, relaxation tends to
keep the sluice in the ground state \cite{Pekola2010}, counteracting LZTs.
In general, determining the explicit dependence of $Q_p$ on $P_{LZ}$ for an
arbitrary environment is a formidable task, which can only be undertaken using
numerical methods.
In order to provide a simple analytical formula to be compared to our data, we
shall further assume that (i) relaxation fully takes place between subsequent
anti-crossings, so that the system approaches every anti-crossing in the ground
state, and (ii) it takes place at the two anti-crossings equally, so that on
average it does not contribute to the pumped charge.
Under (i) and (ii), the problem greatly simplifies, and we find that
$Q_p/2e=1-P_{LZ}$.

Based on this model, in \figref{fig:nonadi}(a) we fit to each experimental
trace the expression
\begin{equation}\label{eq:easyqp}
Q_p(x)= Q_0 \left[1-\exp\left(-2 \pi \lambda x^2 \right) \right]\ ,
\end{equation}
where $x=J_{max}/J_{max}^0$ and $\lambda$ is a
free parameter. The excellent agreement between the data and our single-parameter fit provides strong
evidence that the departure from the adiabatic limit takes place via LZTs, and
that our understanding of decoherence effects, albeit simplified, is essentially correct.

The dependence of $Q_0$ and $\lambda$ on $f$ is shown in
\figref{fig:nonadi}(c,d). Different symbols refer to three different measurement
sets, taken on the same sample using different generators and/or during
different cooldowns. Altogether, they span the frequency range of 40 to 120 MHz.
This range is limited from below by acceptable signal-to-noise ratio, and from
above by the sampling rate of our rf generators. As the three sets yield
consistent results, we discuss them together. The behavior of $Q_0$
[\figref{fig:nonadi}(c)] is not completely clear; it may stem from
frequency-dependent attenuation in the lines. By contrast, $\lambda$
[\figref{fig:nonadi}(d)] displays a clear trend, a steady increase with
decreasing $f$. The solid line in \figref{fig:nonadi}(d) is a best-fit to the
data of the expression $\lambda=f_0/f$, yielding $f_0=\SI{180}{MHz}$.
Our model predicts $f_0^{(th)}=(J^0_{max})^2/(4\hbar \alpha E_C \dng)$, where
$\alpha \approx 4.3$ is a parameter proportional to the steepness of the gate
rise.
Using our estimated $E_C$ and $J^0_{max}$, we obtain $f_0^{(th)} \approx
\SI{120}{MHz}$. We find the agreement between $f_0$ and $f_0^{(th)}$
satisfactory. The discrepancy between the two corresponds to an effective
Josephson energy about 20\% smaller than its nominal value. Among other factors,
this may well be due to influence of noise/decoherence not captured by our
simple model.

\smallskip

Our results demonstrate the feasiblity of using Cooper-pair pumps to investigate
coherent effects in driven quantum systems, and pave the way for further
observations.
By embedding the sluice in a fully superconducting loop (as e.g.~in
Refs.~\onlinecite{Vion2002,Mottonen2008}) and increasing the pumping frequency
up to a few hundred MHz, we expect to achieve coherence times extending over
several pumping periods. This would allow challeging proposals such as
Landau-Zener interferometry with geometric phases \cite{Gasparinetti2011a},
characterization of decoherence induced by an engineered environment
\cite{Solinas2010}, and measurement of the Lamb shift \cite{Gramich2011}, to be
readily implemented.

We would like to thank V. Maisi, M. M\"ott\"onen and O.-P. Saira for useful
discussions, and T. Faivre and M. Meschke for technical assistance. This work
was supported by the European Community FP7 under grants No. 238345 ``GEOMDISS'' and
228464 ``MICROKELVIN'', and by the Finnish National Graduate School in Nanoscience.

%\bibliography{SCP_biblio}

\begin{thebibliography}{32}%
\makeatletter
\providecommand \@ifxundefined [1]{%
 \@ifx{#1\undefined}
}%
\providecommand \@ifnum [1]{%
 \ifnum #1\expandafter \@firstoftwo
 \else \expandafter \@secondoftwo
 \fi
}%
\providecommand \@ifx [1]{%
 \ifx #1\expandafter \@firstoftwo
 \else \expandafter \@secondoftwo
 \fi
}%
\providecommand \natexlab [1]{#1}%
\providecommand \enquote  [1]{``#1''}%
\providecommand \bibnamefont  [1]{#1}%
\providecommand \bibfnamefont [1]{#1}%
\providecommand \citenamefont [1]{#1}%
\providecommand \href@noop [0]{\@secondoftwo}%
\providecommand \href [0]{\begingroup \@sanitize@url \@href}%
\providecommand \@href[1]{\@@startlink{#1}\@@href}%
\providecommand \@@href[1]{\endgroup#1\@@endlink}%
\providecommand \@sanitize@url [0]{\catcode `\\12\catcode `\$12\catcode
  `\&12\catcode `\#12\catcode `\^12\catcode `\_12\catcode `\%12\relax}%
\providecommand \@@startlink[1]{}%
\providecommand \@@endlink[0]{}%
\providecommand \url  [0]{\begingroup\@sanitize@url \@url }%
\providecommand \@url [1]{\endgroup\@href {#1}{\urlprefix }}%
\providecommand \urlprefix  [0]{URL }%
\providecommand \Eprint [0]{\href }%
\@ifxundefined \urlstyle {%
  \providecommand \doi  [0]{\begingroup \@sanitize@url \@doi}%
  \providecommand \@doi [1]{\endgroup \@@startlink {\doibase
  #1}doi:\discretionary {}{}{}#1\@@endlink }%
}{%
  \providecommand \doi  [0]{doi:\discretionary{}{}{}\begingroup
  \urlstyle{rm}\Url }%
}%
\providecommand \doibase [0]{http://dx.doi.org/}%
\providecommand \Doi [0]{\begingroup \@sanitize@url \@Doi }%
\providecommand \@Doi  [1]{\endgroup\@@startlink{\doibase#1}\@@Doi}%
\providecommand \@@Doi [1]{#1\@@endlink}%
\providecommand \selectlanguage [0]{\@gobble}%
\providecommand \bibinfo  [0]{\@secondoftwo}%
\providecommand \bibfield  [0]{\@secondoftwo}%
\providecommand \translation [1]{[#1]}%
\providecommand \BibitemOpen [0]{}%
\providecommand \bibitemStop [0]{}%
\providecommand \bibitemNoStop [0]{.\EOS\space}%
\providecommand \EOS [0]{\spacefactor3000\relax}%
\providecommand \BibitemShut  [1]{\csname bibitem#1\endcsname}%
%</preamble>
\bibitem [{\citenamefont {Blumenthal}\ \emph {et~al.}(2007)\citenamefont
  {Blumenthal}, \citenamefont {Kaestner}, \citenamefont {Li}, \citenamefont
  {Giblin}, \citenamefont {Janssen}, \citenamefont {Pepper}, \citenamefont
  {Anderson}, \citenamefont {Jones},\ and\ \citenamefont
  {Ritchie}}]{Blumenthal2007}%
  \BibitemOpen
  \bibfield  {author} {\bibinfo {author} {\bibfnamefont {M.~D.}\ \bibnamefont
  {Blumenthal}}, \bibinfo {author} {\bibfnamefont {B.}~\bibnamefont
  {Kaestner}}, \bibinfo {author} {\bibfnamefont {L.}~\bibnamefont {Li}},
  \bibinfo {author} {\bibfnamefont {S.~P.}\ \bibnamefont {Giblin}}, \bibinfo
  {author} {\bibfnamefont {T.~J. B.~M.}\ \bibnamefont {Janssen}}, \bibinfo
  {author} {\bibfnamefont {M.}~\bibnamefont {Pepper}}, \bibinfo {author}
  {\bibfnamefont {D.}~\bibnamefont {Anderson}}, \bibinfo {author}
  {\bibfnamefont {G.~A.~C.}\ \bibnamefont {Jones}}, \ and\ \bibinfo {author}
  {\bibfnamefont {D.~A.}\ \bibnamefont {Ritchie}},\ }\href@noop {} {\bibfield
  {journal} {\bibinfo  {journal} {Nat. Phys.},\ }\textbf {\bibinfo {volume}
  {3}},\ \bibinfo {pages} {343} (\bibinfo {year} {2007})}\BibitemShut {NoStop}%
\bibitem [{\citenamefont {Giblin}\ \emph {et~al.}(2010)\citenamefont {Giblin},
  \citenamefont {Wright}, \citenamefont {Fletcher}, \citenamefont {Kataoka},
  \citenamefont {Pepper}, \citenamefont {Janssen}, \citenamefont {Ritchie},
  \citenamefont {Nicoll}, \citenamefont {Anderson},\ and\ \citenamefont
  {Jones}}]{Giblin2010}%
  \BibitemOpen
  \bibfield  {author} {\bibinfo {author} {\bibfnamefont {S.~P.}\ \bibnamefont
  {Giblin}}, \bibinfo {author} {\bibfnamefont {S.~J.}\ \bibnamefont {Wright}},
  \bibinfo {author} {\bibfnamefont {J.~D.}\ \bibnamefont {Fletcher}}, \bibinfo
  {author} {\bibfnamefont {M.}~\bibnamefont {Kataoka}}, \bibinfo {author}
  {\bibfnamefont {M.}~\bibnamefont {Pepper}}, \bibinfo {author} {\bibfnamefont
  {T.~J. B.~M.}\ \bibnamefont {Janssen}}, \bibinfo {author} {\bibfnamefont
  {D.~a.}\ \bibnamefont {Ritchie}}, \bibinfo {author} {\bibfnamefont {C.~a.}\
  \bibnamefont {Nicoll}}, \bibinfo {author} {\bibfnamefont {D.}~\bibnamefont
  {Anderson}}, \ and\ \bibinfo {author} {\bibfnamefont {G.~A.~C.}\ \bibnamefont
  {Jones}},\ }\href@noop {} {\bibfield  {journal} {\bibinfo  {journal} {New
  Journ. of Phys.},\ }\textbf {\bibinfo {volume} {12}},\ \bibinfo {pages}
  {073013} (\bibinfo {year} {2010})}\BibitemShut {NoStop}%
\bibitem [{\citenamefont {Giazotto}\ \emph {et~al.}(2011)\citenamefont
  {Giazotto}, \citenamefont {Spathis}, \citenamefont {Roddaro}, \citenamefont
  {Biswas}, \citenamefont {Taddei}, \citenamefont {Governale},\ and\
  \citenamefont {Sorba}}]{Giazotto2011}%
  \BibitemOpen
  \bibfield  {author} {\bibinfo {author} {\bibfnamefont {F.}~\bibnamefont
  {Giazotto}}, \bibinfo {author} {\bibfnamefont {P.}~\bibnamefont {Spathis}},
  \bibinfo {author} {\bibfnamefont {S.}~\bibnamefont {Roddaro}}, \bibinfo
  {author} {\bibfnamefont {S.}~\bibnamefont {Biswas}}, \bibinfo {author}
  {\bibfnamefont {F.}~\bibnamefont {Taddei}}, \bibinfo {author} {\bibfnamefont
  {M.}~\bibnamefont {Governale}}, \ and\ \bibinfo {author} {\bibfnamefont
  {L.}~\bibnamefont {Sorba}},\ }\href@noop {} {\bibfield  {journal} {\bibinfo
  {journal} {Nat. Phys.},\ }\textbf {\bibinfo {volume} {7}},\ \bibinfo {pages}
  {857} (\bibinfo {year} {2011})}\BibitemShut {NoStop}%
\bibitem [{\citenamefont {Nevou}\ \emph {et~al.}(2011)\citenamefont {Nevou},
  \citenamefont {Liverini}, \citenamefont {Friedli}, \citenamefont
  {Castellano}, \citenamefont {Bismuto}, \citenamefont {Sigg}, \citenamefont
  {Gramm}, \citenamefont {M\"{u}ller},\ and\ \citenamefont
  {Faist}}]{Nevou2011}%
  \BibitemOpen
  \bibfield  {author} {\bibinfo {author} {\bibfnamefont {L.}~\bibnamefont
  {Nevou}}, \bibinfo {author} {\bibfnamefont {V.}~\bibnamefont {Liverini}},
  \bibinfo {author} {\bibfnamefont {P.}~\bibnamefont {Friedli}}, \bibinfo
  {author} {\bibfnamefont {F.}~\bibnamefont {Castellano}}, \bibinfo {author}
  {\bibfnamefont {A.}~\bibnamefont {Bismuto}}, \bibinfo {author} {\bibfnamefont
  {H.}~\bibnamefont {Sigg}}, \bibinfo {author} {\bibfnamefont {F.}~\bibnamefont
  {Gramm}}, \bibinfo {author} {\bibfnamefont {E.}~\bibnamefont {M\"{u}ller}}, \
  and\ \bibinfo {author} {\bibfnamefont {J.}~\bibnamefont {Faist}},\
  }\href@noop {} {\bibfield  {journal} {\bibinfo  {journal} {Nat. Phys.},\
  }\textbf {\bibinfo {volume} {7}},\ \bibinfo {pages} {423} (\bibinfo {year}
  {2011})}\BibitemShut {NoStop}%
\bibitem [{\citenamefont {Pekola}\ \emph {et~al.}(2008)\citenamefont {Pekola},
  \citenamefont {Vartiainen}, \citenamefont {M\"{o}tt\"{o}nen}, \citenamefont
  {Saira}, \citenamefont {Meschke},\ and\ \citenamefont
  {Averin}}]{Pekola2008a}%
  \BibitemOpen
  \bibfield  {author} {\bibinfo {author} {\bibfnamefont {J.~P.}\ \bibnamefont
  {Pekola}}, \bibinfo {author} {\bibfnamefont {J.~J.}\ \bibnamefont
  {Vartiainen}}, \bibinfo {author} {\bibfnamefont {M.}~\bibnamefont
  {M\"{o}tt\"{o}nen}}, \bibinfo {author} {\bibfnamefont {O.-P.}\ \bibnamefont
  {Saira}}, \bibinfo {author} {\bibfnamefont {M.}~\bibnamefont {Meschke}}, \
  and\ \bibinfo {author} {\bibfnamefont {D.~V.}\ \bibnamefont {Averin}},\
  }\href@noop {} {\bibfield  {journal} {\bibinfo  {journal} {Nat. Phys.},\
  }\textbf {\bibinfo {volume} {4}},\ \bibinfo {pages} {120} (\bibinfo {year}
  {2008})}\BibitemShut {NoStop}%
\bibitem [{\citenamefont {F\`{e}ve}\ \emph {et~al.}(2007)\citenamefont
  {F\`{e}ve}, \citenamefont {Mah\'{e}}, \citenamefont {Berroir}, \citenamefont
  {Kontos}, \citenamefont {Pla\c{c}ais}, \citenamefont {Glattli}, \citenamefont
  {Cavanna}, \citenamefont {Etienne},\ and\ \citenamefont {Jin}}]{Feve2007}%
  \BibitemOpen
  \bibfield  {author} {\bibinfo {author} {\bibfnamefont {G.}~\bibnamefont
  {F\`{e}ve}}, \bibinfo {author} {\bibfnamefont {A.}~\bibnamefont {Mah\'{e}}},
  \bibinfo {author} {\bibfnamefont {J.-M.}\ \bibnamefont {Berroir}}, \bibinfo
  {author} {\bibfnamefont {T.}~\bibnamefont {Kontos}}, \bibinfo {author}
  {\bibfnamefont {B.}~\bibnamefont {Pla\c{c}ais}}, \bibinfo {author}
  {\bibfnamefont {D.~C.}\ \bibnamefont {Glattli}}, \bibinfo {author}
  {\bibfnamefont {A.}~\bibnamefont {Cavanna}}, \bibinfo {author} {\bibfnamefont
  {B.}~\bibnamefont {Etienne}}, \ and\ \bibinfo {author} {\bibfnamefont
  {Y.}~\bibnamefont {Jin}},\ }\href@noop {} {\bibfield  {journal} {\bibinfo
  {journal} {Science},\ }\textbf {\bibinfo {volume} {316}},\ \bibinfo {pages}
  {1169} (\bibinfo {year} {2007})}\BibitemShut {NoStop}%
\bibitem [{\citenamefont {Flowers}(2004)}]{Flowers2004}%
  \BibitemOpen
  \bibfield  {author} {\bibinfo {author} {\bibfnamefont {J.}~\bibnamefont
  {Flowers}},\ }\href@noop {} {\bibfield  {journal} {\bibinfo  {journal}
  {Science},\ }\textbf {\bibinfo {volume} {306}},\ \bibinfo {pages} {1324}
  (\bibinfo {year} {2004})}\BibitemShut {NoStop}%
\bibitem [{\citenamefont {Pekola}\ \emph {et~al.}(1999)\citenamefont {Pekola},
  \citenamefont {Toppari}, \citenamefont {Aunola}, \citenamefont {Savolainen},\
  and\ \citenamefont {Averin}}]{Pekola1999}%
  \BibitemOpen
  \bibfield  {author} {\bibinfo {author} {\bibfnamefont {J.~P.}\ \bibnamefont
  {Pekola}}, \bibinfo {author} {\bibfnamefont {J.~J.}\ \bibnamefont {Toppari}},
  \bibinfo {author} {\bibfnamefont {M.}~\bibnamefont {Aunola}}, \bibinfo
  {author} {\bibfnamefont {M.~T.}~\bibnamefont {Savolainen}}, \ and\ \bibinfo
  {author} {\bibfnamefont {D.~V.}\ \bibnamefont {Averin}},\ }\href@noop {}
  {\bibfield  {journal} {\bibinfo  {journal} {Phys. Rev. B},\ }\textbf
  {\bibinfo {volume} {60}},\ \bibinfo {pages} {R9931} (\bibinfo {year}
  {1999})}\BibitemShut {NoStop}%
\bibitem [{\citenamefont {Niskanen}\ \emph {et~al.}(2003)\citenamefont
  {Niskanen}, \citenamefont {Pekola},\ and\ \citenamefont
  {Sepp\"{a}}}]{Niskanen2003}%
  \BibitemOpen
  \bibfield  {author} {\bibinfo {author} {\bibfnamefont {A.~O.}\ \bibnamefont
  {Niskanen}}, \bibinfo {author} {\bibfnamefont {J.~P.}\ \bibnamefont
  {Pekola}}, \ and\ \bibinfo {author} {\bibfnamefont {H.}~\bibnamefont
  {Sepp\"{a}}},\ }\href@noop {} {\bibfield  {journal} {\bibinfo  {journal}
  {Phys. Rev. Lett.},\ }\textbf {\bibinfo {volume} {91}},\ \bibinfo {pages}
  {177003} (\bibinfo {year} {2003})}\BibitemShut {NoStop}%
\bibitem [{\citenamefont {Leone}\ \emph {et~al.}(2008)\citenamefont {Leone},
  \citenamefont {L\'{e}vy},\ and\ \citenamefont {Lafarge}}]{Leone2008a}%
  \BibitemOpen
  \bibfield  {author} {\bibinfo {author} {\bibfnamefont {R.}~\bibnamefont
  {Leone}}, \bibinfo {author} {\bibfnamefont {L.~P.}~\bibnamefont {L\'{e}vy}}, \
  and\ \bibinfo {author} {\bibfnamefont {P.}~\bibnamefont {Lafarge}},\
  }\href@noop {} {\bibfield  {journal} {\bibinfo  {journal} {Phys. Rev.
  Lett.},\ }\textbf {\bibinfo {volume} {100}},\ \bibinfo {pages} {117001}
  (\bibinfo {year} {2008})}\BibitemShut {NoStop}%
\bibitem [{\citenamefont {M\"{o}tt\"{o}nen}\ \emph {et~al.}(2006)\citenamefont
  {M\"{o}tt\"{o}nen}, \citenamefont {Pekola}, \citenamefont {Vartiainen},
  \citenamefont {Brosco},\ and\ \citenamefont {Hekking}}]{Mottonen2006}%
  \BibitemOpen
  \bibfield  {author} {\bibinfo {author} {\bibfnamefont {M.}~\bibnamefont
  {M\"{o}tt\"{o}nen}}, \bibinfo {author} {\bibfnamefont {J.~P.}\ \bibnamefont
  {Pekola}}, \bibinfo {author} {\bibfnamefont {J.~J.}\ \bibnamefont
  {Vartiainen}}, \bibinfo {author} {\bibfnamefont {V.}~\bibnamefont {Brosco}},
  \ and\ \bibinfo {author} {\bibfnamefont {F.~W.~J.}\ \bibnamefont {Hekking}},\
  }\href@noop {} {\bibfield  {journal} {\bibinfo  {journal} {Phys. Rev. B},\
  }\textbf {\bibinfo {volume} {73}},\ \bibinfo {pages} {214523} (\bibinfo
  {year} {2006})}\BibitemShut {NoStop}%
\bibitem [{\citenamefont {Aunola}\ and\ \citenamefont
  {Toppari}(2003)}]{Aunola2003}%
  \BibitemOpen
  \bibfield  {author} {\bibinfo {author} {\bibfnamefont {M.}~\bibnamefont
  {Aunola}}\ and\ \bibinfo {author} {\bibfnamefont {J.~J.}\ \bibnamefont
  {Toppari}},\ }\href@noop {} {\bibfield  {journal} {\bibinfo  {journal} {Phys.
  Rev. B},\ }\textbf {\bibinfo {volume} {68}},\ \bibinfo {pages} {020502(R)}
  (\bibinfo {year} {2003})}\BibitemShut {NoStop}%
\bibitem [{\citenamefont {M\"{o}tt\"{o}nen}\ \emph {et~al.}(2008)\citenamefont
  {M\"{o}tt\"{o}nen}, \citenamefont {Vartiainen},\ and\ \citenamefont
  {Pekola}}]{Mottonen2008}%
  \BibitemOpen
  \bibfield  {author} {\bibinfo {author} {\bibfnamefont {M.}~\bibnamefont
  {M\"{o}tt\"{o}nen}}, \bibinfo {author} {\bibfnamefont {J.~J.}\ \bibnamefont
  {Vartiainen}}, \ and\ \bibinfo {author} {\bibfnamefont {J.~P.}\ \bibnamefont
  {Pekola}},\ }\href@noop {} {\bibfield  {journal} {\bibinfo  {journal} {Phys.
  Rev. Lett.},\ }\textbf {\bibinfo {volume} {100}},\ \bibinfo {pages} {177201}
  (\bibinfo {year} {2008})}\BibitemShut {NoStop}%
\bibitem [{\citenamefont {Gasparinetti}\ \emph {et~al.}(2011)\citenamefont
  {Gasparinetti}, \citenamefont {Solinas},\ and\ \citenamefont
  {Pekola}}]{Gasparinetti2011a}%
  \BibitemOpen
  \bibfield  {author} {\bibinfo {author} {\bibfnamefont {S.}~\bibnamefont
  {Gasparinetti}}, \bibinfo {author} {\bibfnamefont {P.}~\bibnamefont
  {Solinas}}, \ and\ \bibinfo {author} {\bibfnamefont {J.~P.}\ \bibnamefont
  {Pekola}},\ }\href@noop {} {\bibfield  {journal} {\bibinfo  {journal} {Phys.
  Rev. Lett.},\ }\textbf {\bibinfo {volume} {107}},\ \bibinfo {pages} {207002}
  (\bibinfo {year} {2011})}\BibitemShut {NoStop}%
\bibitem [{\citenamefont {Niskanen}\ \emph {et~al.}(2005)\citenamefont
  {Niskanen}, \citenamefont {Kivioja}, \citenamefont {Sepp\"{a}},\ and\
  \citenamefont {Pekola}}]{Niskanen2005}%
  \BibitemOpen
  \bibfield  {author} {\bibinfo {author} {\bibfnamefont {A.~O.}\ \bibnamefont
  {Niskanen}}, \bibinfo {author} {\bibfnamefont {J.~M.}\ \bibnamefont
  {Kivioja}}, \bibinfo {author} {\bibfnamefont {H.}~\bibnamefont {Sepp\"{a}}},
  \ and\ \bibinfo {author} {\bibfnamefont {J.~P.}\ \bibnamefont {Pekola}},\
  }\href@noop {} {\bibfield  {journal} {\bibinfo  {journal} {Phys. Rev. B},\
  }\textbf {\bibinfo {volume} {71}},\ \bibinfo {pages} {012513} (\bibinfo {year}
  {2005})}\BibitemShut {NoStop}%
\bibitem [{\citenamefont {Pekola}\ \emph {et~al.}(2010)\citenamefont {Pekola},
  \citenamefont {Brosco}, \citenamefont {M\"{o}tt\"{o}nen}, \citenamefont
  {Solinas},\ and\ \citenamefont {Shnirman}}]{Pekola2010}%
  \BibitemOpen
  \bibfield  {author} {\bibinfo {author} {\bibfnamefont {J.~P.}\ \bibnamefont
  {Pekola}}, \bibinfo {author} {\bibfnamefont {V.}~\bibnamefont {Brosco}},
  \bibinfo {author} {\bibfnamefont {M.}~\bibnamefont {M\"{o}tt\"{o}nen}},
  \bibinfo {author} {\bibfnamefont {P.}~\bibnamefont {Solinas}}, \ and\
  \bibinfo {author} {\bibfnamefont {A.}~\bibnamefont {Shnirman}},\ }\href@noop
  {} {\bibfield  {journal} {\bibinfo  {journal} {Phys. Rev. Lett.},\ }\textbf
  {\bibinfo {volume} {105}},\ \bibinfo {pages} {030401} (\bibinfo {year}
  {2010})}\BibitemShut {NoStop}%
\bibitem [{\citenamefont {Vartiainen}\ \emph {et~al.}(2007)\citenamefont
  {Vartiainen}, \citenamefont {M\"{o}tt\"{o}nen}, \citenamefont {Pekola},\ and\
  \citenamefont {Kemppinen}}]{Vartiainen2007}%
  \BibitemOpen
  \bibfield  {author} {\bibinfo {author} {\bibfnamefont {J.~J.}\ \bibnamefont
  {Vartiainen}}, \bibinfo {author} {\bibfnamefont {M.}~\bibnamefont
  {M\"{o}tt\"{o}nen}}, \bibinfo {author} {\bibfnamefont {J.~P.}\ \bibnamefont
  {Pekola}}, \ and\ \bibinfo {author} {\bibfnamefont {A.}~\bibnamefont
  {Kemppinen}},\ }\href@noop {} {\bibfield  {journal} {\bibinfo  {journal}
  {Appl. Phys. Lett.},\ }\textbf {\bibinfo {volume} {90}},\ \bibinfo {pages}
  {082102} (\bibinfo {year} {2007})}\BibitemShut {NoStop}%
\bibitem [{\citenamefont {Joyez}\ \emph {et~al.}(1994)\citenamefont {Joyez},
  \citenamefont {Lafarge}, \citenamefont {Filipe}, \citenamefont {Esteve},\
  and\ \citenamefont {Devoret}}]{Joyez1994}%
  \BibitemOpen
  \bibfield  {author} {\bibinfo {author} {\bibfnamefont {P.}~\bibnamefont
  {Joyez}}, \bibinfo {author} {\bibfnamefont {P.}~\bibnamefont {Lafarge}},
  \bibinfo {author} {\bibfnamefont {A.}~\bibnamefont {Filipe}}, \bibinfo
  {author} {\bibfnamefont {D.}~\bibnamefont {Esteve}}, \ and\ \bibinfo {author}
  {\bibfnamefont {M.~H.}\ \bibnamefont {Devoret}},\ }\href@noop {} {\bibfield
  {journal} {\bibinfo  {journal} {Phys. Rev. Lett.},\ }\textbf {\bibinfo
  {volume} {72}},\ \bibinfo {pages} {2458} (\bibinfo {year}
  {1994})}\BibitemShut {NoStop}%
\bibitem [{\citenamefont {Aumentado}\ \emph {et~al.}(2004)\citenamefont
  {Aumentado}, \citenamefont {Keller}, \citenamefont {Martinis},\ and\
  \citenamefont {Devoret}}]{Aumentado2004}%
  \BibitemOpen
  \bibfield  {author} {\bibinfo {author} {\bibfnamefont {J.}~\bibnamefont
  {Aumentado}}, \bibinfo {author} {\bibfnamefont {M.~W.}~\bibnamefont {Keller}},
  \bibinfo {author} {\bibfnamefont {J.~M.}~\bibnamefont {Martinis}}, \ and\
  \bibinfo {author} {\bibfnamefont {M.~H.}\ \bibnamefont {Devoret}},\
  }\href@noop {} {\bibfield  {journal} {\bibinfo  {journal} {Phys. Rev.
  Lett.},\ }\textbf {\bibinfo {volume} {92}},\ \bibinfo {pages} {066802}
  (\bibinfo {year} {2004})}\BibitemShut {NoStop}%
\bibitem [{\citenamefont {Naaman}\ and\ \citenamefont
  {Aumentado}(2006)}]{Naaman2006}%
  \BibitemOpen
  \bibfield  {author} {\bibinfo {author} {\bibfnamefont {O.}~\bibnamefont
  {Naaman}}\ and\ \bibinfo {author} {\bibfnamefont {J.}~\bibnamefont
  {Aumentado}},\ }\href@noop {} {\bibfield  {journal} {\bibinfo  {journal}
  {Phys. Rev. B},\ }\textbf {\bibinfo {volume} {73}},\ \bibinfo {pages}
  {172504} (\bibinfo {year} {2006})}\BibitemShut {NoStop}%
\bibitem [{\citenamefont {Ferguson}\ \emph {et~al.}(2006)\citenamefont
  {Ferguson}, \citenamefont {Court}, \citenamefont {Hudson},\ and\
  \citenamefont {Clark}}]{Ferguson2006}%
  \BibitemOpen
  \bibfield  {author} {\bibinfo {author} {\bibfnamefont {A.~J.}~\bibnamefont
  {Ferguson}}, \bibinfo {author} {\bibfnamefont {N.~A.}~\bibnamefont {Court}},
  \bibinfo {author} {\bibfnamefont {F.~E.}~\bibnamefont {Hudson}}, \ and\
  \bibinfo {author} {\bibfnamefont {R.~G.}~\bibnamefont {Clark}},\ }\href@noop
  {} {\bibfield  {journal} {\bibinfo  {journal} {Phys. Rev. Lett.},\ }\textbf
  {\bibinfo {volume} {97}},\ \bibinfo {pages} {106603} (\bibinfo {year}
  {2006})}\BibitemShut {NoStop}%
\bibitem [{\citenamefont {Shaw}\ \emph {et~al.}(2008)\citenamefont {Shaw},
  \citenamefont {Lutchyn}, \citenamefont {Delsing},\ and\ \citenamefont
  {Echternach}}]{Shaw2008}%
  \BibitemOpen
  \bibfield  {author} {\bibinfo {author} {\bibfnamefont {M.~D.}~\bibnamefont
  {Shaw}}, \bibinfo {author} {\bibfnamefont {R.~M.}~\bibnamefont {Lutchyn}},
  \bibinfo {author} {\bibfnamefont {P.}~\bibnamefont {Delsing}}, \ and\
  \bibinfo {author} {\bibfnamefont {P.~M.}~\bibnamefont {Echternach}},\
  }\href@noop {} {\bibfield  {journal} {\bibinfo  {journal} {Phys. Rev. B},\
  }\textbf {\bibinfo {volume} {78}},\ \bibinfo {pages} {024503} (\bibinfo
  {year} {2008})}\BibitemShut {NoStop}%
\bibitem [{\citenamefont {Persson}\ \emph {et~al.}(2010)\citenamefont
  {Persson}, \citenamefont {Wilson}, \citenamefont {Sandberg},\ and\
  \citenamefont {Delsing}}]{Persson2010}%
  \BibitemOpen
  \bibfield  {author} {\bibinfo {author} {\bibfnamefont {F.}~\bibnamefont
  {Persson}}, \bibinfo {author} {\bibfnamefont {C.~M.}\ \bibnamefont {Wilson}},
  \bibinfo {author} {\bibfnamefont {M.}~\bibnamefont {Sandberg}}, \ and\
  \bibinfo {author} {\bibfnamefont {P.}~\bibnamefont {Delsing}},\ }\href@noop
  {} {\bibfield  {journal} {\bibinfo  {journal} {Phys. Rev. B},\ }\textbf
  {\bibinfo {volume} {82}},\ \bibinfo {pages} {134533} (\bibinfo {year}
  {2010})}\BibitemShut {NoStop}%
\bibitem [{\citenamefont {Barends}\ \emph {et~al.}(2011)\citenamefont
  {Barends}, \citenamefont {Wenner}, \citenamefont {Lenander}, \citenamefont
  {Chen}, \citenamefont {Bialczak}, \citenamefont {Kelly}, \citenamefont
  {Lucero}, \citenamefont {O'Malley}, \citenamefont {Mariantoni}, \citenamefont
  {Sank}, \citenamefont {Wang}, \citenamefont {White}, \citenamefont {Yin},
  \citenamefont {Zhao}, \citenamefont {Cleland}, \citenamefont {Martinis},\
  and\ \citenamefont {Baselmans}}]{Barends2011}%
  \BibitemOpen
  \bibfield  {author} {\bibinfo {author} {\bibfnamefont {R.}~\bibnamefont
  {Barends}}, \bibinfo {author} {\bibfnamefont {J.}~\bibnamefont {Wenner}},
  \bibinfo {author} {\bibfnamefont {M.}~\bibnamefont {Lenander}}, \bibinfo
  {author} {\bibfnamefont {Y.}~\bibnamefont {Chen}}, \bibinfo {author}
  {\bibfnamefont {R.~C.}\ \bibnamefont {Bialczak}}, \bibinfo {author}
  {\bibfnamefont {J.}~\bibnamefont {Kelly}}, \bibinfo {author} {\bibfnamefont
  {E.}~\bibnamefont {Lucero}}, \bibinfo {author} {\bibfnamefont
  {P.}~\bibnamefont {O'Malley}}, \bibinfo {author} {\bibfnamefont
  {M.}~\bibnamefont {Mariantoni}}, \bibinfo {author} {\bibfnamefont
  {D.}~\bibnamefont {Sank}}, \bibinfo {author} {\bibfnamefont {H.}~\bibnamefont
  {Wang}}, \bibinfo {author} {\bibfnamefont {T.~C.}\ \bibnamefont {White}},
  \bibinfo {author} {\bibfnamefont {Y.}~\bibnamefont {Yin}}, \bibinfo {author}
  {\bibfnamefont {J.}~\bibnamefont {Zhao}}, \bibinfo {author} {\bibfnamefont
  {A.~N.}\ \bibnamefont {Cleland}}, \bibinfo {author} {\bibfnamefont {J.~M.}\
  \bibnamefont {Martinis}}, \ and\ \bibinfo {author} {\bibfnamefont {J.~J.~A.}\
  \bibnamefont {Baselmans}},\ }\href@noop {} {\bibfield  {journal} {\bibinfo
  {journal} {Appl. Phys. Lett.},\ }\textbf {\bibinfo {volume} {99}},\ \bibinfo
  {pages} {113507} (\bibinfo {year} {2011})}\BibitemShut {NoStop}%
\bibitem [{\citenamefont {Saira}\ \emph {et~al.}(2012)\citenamefont {Saira},
  \citenamefont {Kemppinen}, \citenamefont {Maisi},\ and\ \citenamefont
  {Pekola}}]{Saira2012}%
  \BibitemOpen
  \bibfield  {author} {\bibinfo {author} {\bibfnamefont {O.-P.}\ \bibnamefont
  {Saira}}, \bibinfo {author} {\bibfnamefont {A.}~\bibnamefont {Kemppinen}},
  \bibinfo {author} {\bibfnamefont {V.~F.}~\bibnamefont {Maisi}}, \ and\
  \bibinfo {author} {\bibfnamefont {J.~P.}\ \bibnamefont {Pekola}},\ }\href@noop {}
  {\bibfield  {journal} {\bibinfo  {journal} {Phys. Rev. B},\ }\textbf
  {\bibinfo {volume} {85}},\ \bibinfo {pages} {012504} (\bibinfo {year}
  {2012})}\BibitemShut {NoStop}%
\bibitem [{not()}]{note_qptunnel}%
  \BibitemOpen
  \href@noop {} {}\bibinfo {note} {$\Delta E^\pm [n,n_g] = 4E_C
  [(n-n_g)^2-(n-n_g \mp 1/2)^2]$, where $n$ is the number of excess
  Cooper-pairs on the island.}\BibitemShut {Stop}%
\bibitem [{\citenamefont {Fulton}\ \emph {et~al.}(1989)\citenamefont {Fulton},
  \citenamefont {Gammel}, \citenamefont {Bishop},\ and\ \citenamefont
  {Dunkleberger}}]{Fulton1989}%
  \BibitemOpen
  \bibfield  {author} {\bibinfo {author} {\bibfnamefont {T.~A.}\ \bibnamefont
  {Fulton}}, \bibinfo {author} {\bibfnamefont {P.~L.}\ \bibnamefont {Gammel}},
  \bibinfo {author} {\bibfnamefont {D.~J.}\ \bibnamefont {Bishop}},
  \bibinfo {author} {\bibfnamefont {L.~N.}\ \bibnamefont {Dunkleberger}}, \ and\
  \bibinfo {author} {\bibfnamefont {G.~J.}\ \bibnamefont {Dolan}},\
   }\href@noop {} {\bibfield  {journal} {\bibinfo  {journal} {Phys. Rev.
  Lett.},\ }\textbf {\bibinfo {volume} {63}},\ \bibinfo {pages} {1307}
  (\bibinfo {year} {1989})}\BibitemShut {NoStop}%
\bibitem [{\citenamefont {Solinas}\ \emph {et~al.}(2010)\citenamefont
  {Solinas}, \citenamefont {M\"{o}tt\"{o}nen}, \citenamefont {Salmilehto},\
  and\ \citenamefont {Pekola}}]{Solinas2010}%
  \BibitemOpen
  \bibfield  {author} {\bibinfo {author} {\bibfnamefont {P.}~\bibnamefont
  {Solinas}}, \bibinfo {author} {\bibfnamefont {M.}~\bibnamefont
  {M\"{o}tt\"{o}nen}}, \bibinfo {author} {\bibfnamefont {J.}~\bibnamefont
  {Salmilehto}}, \ and\ \bibinfo {author} {\bibfnamefont {J.~P.}\ \bibnamefont
  {Pekola}},\ }\href@noop {} {\bibfield  {journal} {\bibinfo  {journal} {Phys.
  Rev. B},\ }\textbf {\bibinfo {volume} {82}},\ \bibinfo {pages} {134517}
  (\bibinfo {year} {2010})}\BibitemShut {NoStop}%
\bibitem [{\citenamefont {Russomanno}\ \emph {et~al.}(2011)\citenamefont
  {Russomanno}, \citenamefont {Pugnetti}, \citenamefont {Brosco},\ and\
  \citenamefont {Fazio}}]{Russomanno2011}%
  \BibitemOpen
  \bibfield  {author} {\bibinfo {author} {\bibfnamefont {A.}~\bibnamefont
  {Russomanno}}, \bibinfo {author} {\bibfnamefont {S.}~\bibnamefont
  {Pugnetti}}, \bibinfo {author} {\bibfnamefont {V.}~\bibnamefont {Brosco}}, \
  and\ \bibinfo {author} {\bibfnamefont {R.}~\bibnamefont {Fazio}},\
  }\href@noop {} {\bibfield  {journal} {\bibinfo  {journal} {Phys. Rev. B},\
  }\textbf {\bibinfo {volume} {83}},\ \bibinfo {pages} {214508} (\bibinfo
  {year} {2011})}\BibitemShut {NoStop}%
\bibitem [{\citenamefont {Shevchenko}\ \emph {et~al.}(2010)\citenamefont
  {Shevchenko}, \citenamefont {Ashhab},\ and\ \citenamefont
  {Nori}}]{Shevchenko2010}%
  \BibitemOpen
  \bibfield  {author} {\bibinfo {author} {\bibfnamefont {S.~N.}\ \bibnamefont
  {Shevchenko}}, \bibinfo {author} {\bibfnamefont {S.}~\bibnamefont {Ashhab}},
  \ and\ \bibinfo {author} {\bibfnamefont {F.}~\bibnamefont {Nori}},\
  }\href@noop {} {\bibfield  {journal} {\bibinfo  {journal} {Phys. Rep.},\
  }\textbf {\bibinfo {volume} {492}},\ \bibinfo {pages} {1} (\bibinfo {year}
  {2010})}\BibitemShut {NoStop}%
\bibitem [{\citenamefont {Vion}\ \emph {et~al.}(2002)\citenamefont {Vion},
  \citenamefont {Aassime}, \citenamefont {Cottet}, \citenamefont {Joyez},
  \citenamefont {Pothier}, \citenamefont {Urbina}, \citenamefont {Esteve},\
  and\ \citenamefont {Devoret}}]{Vion2002}%
  \BibitemOpen
  \bibfield  {author} {\bibinfo {author} {\bibfnamefont {D.}~\bibnamefont
  {Vion}}, \bibinfo {author} {\bibfnamefont {A.}~\bibnamefont {Aassime}},
  \bibinfo {author} {\bibfnamefont {A.}~\bibnamefont {Cottet}}, \bibinfo
  {author} {\bibfnamefont {P.}~\bibnamefont {Joyez}}, \bibinfo {author}
  {\bibfnamefont {H.}~\bibnamefont {Pothier}}, \bibinfo {author} {\bibfnamefont
  {C.}~\bibnamefont {Urbina}}, \bibinfo {author} {\bibfnamefont
  {D.}~\bibnamefont {Esteve}}, \ and\ \bibinfo {author} {\bibfnamefont {M.~H.}\
  \bibnamefont {Devoret}},\ }\href@noop {} {\bibfield  {journal} {\bibinfo
  {journal} {Science},\ }\textbf {\bibinfo {volume} {296}},\ \bibinfo {pages}
  {886} (\bibinfo {year} {2002})}\BibitemShut {NoStop}%
\bibitem [{\citenamefont {Gramich}\ \emph {et~al.}(2011)\citenamefont
  {Gramich}, \citenamefont {Solinas}, \citenamefont {M\"{o}tt\"{o}nen},
  \citenamefont {Pekola},\ and\ \citenamefont {Ankerhold}}]{Gramich2011}%
  \BibitemOpen
  \bibfield  {author} {\bibinfo {author} {\bibfnamefont {V.}~\bibnamefont
  {Gramich}}, \bibinfo {author} {\bibfnamefont {P.}~\bibnamefont {Solinas}},
  \bibinfo {author} {\bibfnamefont {M.}~\bibnamefont {M\"{o}tt\"{o}nen}},
  \bibinfo {author} {\bibfnamefont {J.~P.}\ \bibnamefont {Pekola}}, \ and\
  \bibinfo {author} {\bibfnamefont {J.}~\bibnamefont {Ankerhold}},\ }\href@noop
  {} {\bibfield  {journal} {\bibinfo  {journal} {Phys. Rev. A},\ }\textbf
  {\bibinfo {volume} {84}},\ \bibinfo {pages} {052103} (\bibinfo {year}
  {2011})}\BibitemShut {NoStop}%
\end{thebibliography}

%

\end{document}